# Compression Aware Physical Database Design

Hideaki Kimura*
Brown University
Providence, RI, USA
hkimura@cs.brown.edu

Vivek Narasayya
Microsoft Research
Redmond, WA, USA
viveknar@microsoft.com

Manoj Syamala
Microsoft Research
Redmond, WA, USA
manojsy@microsoft.com

## ABSTRACT
Modern RDBMSs support the ability to compress data using methods such as null suppression and dictionary encoding. Data compression offers the promise of significantly reducing storage requirements and improving I/O performance for decision support queries. However, compression can also slow down update and query performance due to the CPU costs of compression and decompression. In this paper, we study how data compression affects choice of appropriate physical database design, such as indexes, for a given workload. We observe that approaches that decouple the decision of whether or not to choose an index from whether or not to compress the index can result in poor solutions. Thus, we focus on the novel problem of integrating compression into physical database design in a scalable manner. We have implemented our techniques by modifying Microsoft SQL Server and the Database Engine Tuning Advisor (DTA) physical design tool. Our techniques are general and are potentially applicable to DBMSs that support other compression methods. Our experimental results on real world as well as TPC-H benchmark workloads demonstrate the effectiveness of our techniques.

## 1. Introduction
Relational database systems (RDBMSs) today support lossless data compression methods such as null suppression and dictionary encoding [5] [14] [13] on physical design structures such as heaps, clustered and non-clustered indexes. Depending on the compression method and the distribution of values in the columns of the index, a compressed index sometimes can require only a small fraction of the storage space of an uncompressed index. For decision support queries which often scan large indexes, compression can result in significantly reduced I/O costs [12]. While compression can improve performance, it also has the potential to slow down performance significantly. In most RDBMSs today, processing a query requires decompressing the data, which incurs significant CPU costs. This can slow down queries that are already CPU bound. Likewise, updates (INSERT/UPDATE statements) also require additional CPU costs since the updated data must be compressed. Thus, compression introduces a potentially significant new dimension to the physical database design problem.

The problem of determining a good physical database design for a complex query workload is an important and challenging problem for database administrators (DBAs). There has been work in the research community as well as industry to automate the process of physical database design (e.g. [7] [4] [15]). In fact, most RDBMSs today support automated physical design tools that assist DBAs in making judicious physical design choices. Such tools typically take as input a *workload* of SQL query and update statements and a *storage bound*, and produce a *configuration* (i.e. set of indexes) that optimizes workload performance, while not exceeding the given storage bound. The performance metric that these tools try to optimize is the query optimizer's total *estimated costs* of statements in workload. To the best of our knowledge however, none of the prior work on physical database design takes into account the impact of data compression.

In this paper, we study the problem of how to effectively incorporate compression into automated physical database design. We focus primarily on indexes and briefly discuss how our techniques extend to other physical design structures such as partial indexes and materialized views (which can also be compressed in today's RDBMSs). An important observation that motivates this work is that decoupling the decision of whether or not to choose an index from whether or not to compress the index can result in poor solutions. Intuitively this is because different indexes achieve different *compression fractions* (i.e. ratio of compressed size to uncompressed size), and therefore the I/O reduction as well as the update cost of an index for a query/update relative to another index can change significantly once compression is considered. For example, consider a simple strategy of staging index selection and compression; i.e. select indexes without considering compression, compress the selected indexes, and repeat the process if the space consumed is below the storage bound. The following example illustrates why the staged approach can result in poor solutions.

*Example 1*. Consider a table *Sales* (*OrderID*, *Shipdate*, *State*, *Price*, *Discount*,…) and a query $Q_1$ = **SELECT SUM(***Price* * *Discount***) FROM** *Sales* **WHERE** *Shipdate* **BETWEEN '01-01-2009' and '12-31-2009' AND** *State* = **'CA'**. Let index $I_1$ = (*Shipdate*, *State*) and $I_2$ = (*Shipdate*, *State*, *Price*, *Discount*) be two indexes on *Sales*. Suppose the given storage bound is 100 GB and the sizes of indexes $I_1$, $I_2$ respectively are 95 GB, 170 GB. Let $I^C_1$ and $I^C_2$ be the *compressed* versions of $I_1$ and $I_2$ respectively and let the sizes of $I^C_1$ and $I^C_2$ respectively be 50 GB and 90 GB. Observe that if we select indexes without considering compression, then we would pick $I_1$, since $I_2$ does not fit within the given space budget. Once $I_1$ is picked, there will not be enough storage to add $I^C_2$ later. On the other hand if we consider compression during the index selection process, we would have picked $I^C_2$ whose size is below the given storage bound. $I^C_2$ is a covering index for $Q_1$ (i.e. it contains all columns required to answer $Q_1$) and thus can improve the query's I/O performance significantly.

Similarly, choosing an index without considering how its CPU overhead will increase if the index is subsequently compressed can also result in poor solutions illustrated in the example below.

---

*Example 2*. Consider a *covering* index $I_3$=(*Shipdate, State, Price, Discount*) on Sales for $Q_1$. $I_3$ significantly speeds $Q_1$ up and is likely to chosen if there is enough storage. However, compressing $I_3$ results in high CPU overheads to compress (during updates) and decompress (during $Q_1$) its data pages. Due to the CPU overheads, an anecdotal outcome of blindly compressing every suggested index is a lower database throughput with a larger storage bound especially when the workload is update-intensive.

We note that the above observation on the importance of integrating compression into physical database design is borne out in our empirical evaluations as well.

The need to integrate compression into physical database design leads to several novel technical challenges, which we study in this paper. First, a large number of new (compressed) indexes must be considered. In principle, for each index, *compressed variants* of that index must also be considered, one per compression method available in the RDBMS. For example, in Microsoft SQL Server both null suppression and dictionary encoding methods are available for compressing an index. For each compressed index, we need to accurately and efficiently *estimate the size* (i.e. number of pages) of each index, since this information is crucial for the query optimizer in determining the cost of the execution plan that uses the index. Observe that, for an uncompressed index, it is relatively straightforward to estimate the size once the number of rows and average row length is known. However, for a compressed index, the size can depend crucially on the compression *method* and the value distribution of columns in the index. An index that is dictionary compressed can have a very different size than if compressed using null suppression. Sampling has been proposed as a mechanism for speeding up size estimation of compressed indexes, i.e. a sample is obtained and the index is created on the sample. The compression fraction thus obtained is used to infer the size of the full compressed index. For example, [11] studies the accuracy of using sampling for estimating size of indexes compressed using null suppression and dictionary encoding. Although sampling results in sufficiently accurate size estimates in practice, the key challenge is performance since most of the time is spent in *creating the index* on the sample. Indeed, as we show in this paper, without additional optimizations, the performance of physical design tools would be unacceptable. Thus, we develop a new *index size estimation framework* that can significantly reduce the overhead to create indexes on the sample while still maintaining the desired level of accuracy.

Second, compression greatly amplifies the *space vs. time trade-off* that physical design tools must deal with. For example, for scalability reasons, today's physical design tools are architected to perform early pruning by eliminating indexes that are not part of the "best" configuration(s) for at least one query in the workload. Such pruning is typically done based purely on query costs. Thus a compressed index that reduces storage space significantly while only increasing query costs a little will likely be pruned. However, retaining such indexes can improve the overall quality of solutions noticeably since the reduced storage allows other indexes to be added (potentially benefiting many other queries). We propose principled adaptations to algorithms used in today's physical design tools to better handle the amplification of space vs. time tradeoff due to compression.

Third, physical design tools today rely on extensions to the query optimizer API to support "what-if" analysis: given a configuration and a query, this API returns the *optimizer's estimated cost* of the query under the (hypothetical) configuration. To integrate compression into physical design also required extending the query optimizer's *cost model* to reflect the cost of using a compressed index. We have extended the cost model of Microsoft SQL Server 2008 R2 to make it "compression-aware". Our cost model captures CPU costs of compression and decompression as well as I/O cost reduction due to compression.

We have implemented the techniques described above in Microsoft SQL Server's automated physical database design tool: Database Engine Tuning Advisor (DTA) [3] so that it can recommend a combination of compressed and uncompressed indexes. Experimental results on the TPC-H benchmark workload as well as on a real-world customer workload demonstrate the effectiveness of our techniques. In the following sections, we first briefly review compression in database systems and then describe the details of our techniques.

## 2. Background
### 2.1 Compression Methods in Databases
The database community has studied several compression techniques in the context of query processing. Among the various compression methods, virtually all modern DBMSs provide dictionary encoding and NULL/prefix suppression [5] [14] [13] because they are relatively easy to implement and well suited for query processing.

Dictionary encoding compresses a given data page by finding frequently occurring values and replacing them with small pointers to a *dictionary*, which contains the distinct set of replaced values. For example, a data page which contains the values {AA, BB, BB, AA} will be compressed to a dictionary {AA=1, BB=2} and a compressed data page {1, 2, 2, 1}. Some databases (e.g. IBM DB2) maintain one dictionary across all data pages in a table partition (*global dictionary*) while other databases (e.g., Oracle) maintain one dictionary per disk block (*local dictionary*). In general, global dictionary achieves better compression while local dictionary provides greater flexibility and better update performance.

NULL suppression eliminates leading NULLs or blank spaces. Typically, databases replace them with a special character and a length of the sequence of NULLs or spaces. For example, a fixed length CHAR value with many leading NULLs "00000abc" will be replaced to "@5abc" where "@" is the special character to represent compressed NULLs. Prefix suppression is similar to NULL suppression, but it compresses arbitrary prefix instead of NULLs. For example, the values {aaabc, aaacd, aaade} share the leading prefix "aaa". Prefix compression replaces them with {@bc, @cd, @de} where "@" represents the leading "aaa".

Microsoft SQL Server supports NULL suppression, prefix suppression and local dictionary compression. More details of these compression schemes can be found in [13] [10].

### 2.2 Estimating Compression Fraction
Most benefits of data compression are due to the reduced data size. Thus, accurately estimating the size of a compressed index, or equivalently the compression fraction (*CF*) is important. CF is defined as the ratio of the size of the compressed index to the size of the uncompressed index. Note that the compression fraction depends on the compression method used. The option of scanning the entire data and running the compression method on it will yield an accurate estimate of the compression fraction of the index but is prohibitively expensive on large databases. Another approach is to estimate the compression fraction based only on statistics of columns in the index (e.g. histograms or the number of distinct values). Such statistics are typically maintained by the query optimizer for purposes of cardinality estimation. For example, in [5] the authors develop an analytical *Compression Estimator* to estimate the fraction for delta RID compression and prefix suppression using those statistics. However, such a static approach has to assume uniform distribution (or worst-case distribution as assumed in the paper) and also requires index-specific statistics (e.g., cluster ratio). Collecting such statistics for each index is expensive unless the index to be compressed already exists in the database.

Another approach is using random sampling. In [11] the authors analyze the accuracy of a sampling based estimation method for the compression fraction (called *SampleCF*). *SampleCF*(*I*) for an index *I* works as follows. It first takes a random sample of the data using a given sampling fraction *f* (e.g. a 1% sample) and creates the index *I* on the sample (say the index size is S). It then compresses the index using the given compression method to obtain the compressed index $I^c$ (say the index size is $S^c$). *SampleCF* then returns the compression fraction as $S^c/S$. The advantage of *SampleCF* method is that it works for every compression method and is agnostic to its implementation. The results in [11] show that *SampleCF* can be quite accurate for NULL suppression, prefix suppression and global dictionary compression. However, the main drawback of *SampleCF* is that, although it is much more efficient than building an index on the full data, it is still expensive to: (a) Take a uniform random sample from the original table for each invocation of *SampleCF*. (b) Create an index on a sample (due to the cost of sorting and compression).

## 3. Solution Overview

We have incorporated the techniques presented in this paper for compression aware physical database design into Microsoft SQL Server's tool Database Engine Tuning Advisor (DTA). The architecture of this tool along with highlights of extensions we made to handle compression is shown in Figure 1. We take as input a workload of SQL statements and a storage bound and produce as output a physical design recommendation consisting of compressed and uncompressed physical design structures (indexes and materialized views).

Today's physical design tools such as DTA rely on the ability to perform *what-if analysis*, i.e. request the query optimizer to return a plan for a given query and a given hypothetical physical design configuration. In order to deal with compressed indexes and materialized views, we had to extend the optimizer's cost model to make it compression aware, i.e. handle compressed indexes in the configuration. Our new compression-aware cost model (described in **Appendix A**) considers the CPU costs to compress and decompress data in compressed indexes.

As described in the introduction, a key new challenge that arises is accurately and efficiently *estimating the size* of compressed indexes considered by the tool. As confirmed in our empirical evaluation (Section 7.1), the scalability of physical design tools crucially depends on addressing this challenge. We use the sampling based method described in Section 2.2 (*SampleCF*), but also develop faster alternative methods based on *deducing* the size without need for sorting and compressing samples (Section 4). In Section 5 we show how given a set of indexes whose compressed sizes need to be estimated, we can do that efficiently (using a combination of *SampleCF* and deducing compressed sizes of others) while still maintaining a desired level of accuracy.

Physical design tools must work with a given storage bound, i.e. a space budget. Thus, they need to deal with the space vs. performance trade-off. However, with compression, this trade-off is greatly amplified. A *compressed* index although sub-optimal for a particular query compared to the uncompressed index, may save a lot of space thereby allowing other indexes to benefit the same or other queries. We propose and evaluate principled techniques for addressing this space-time tradeoff that can be applied to today's physical design tools. In the context of DTA, this affects the *Candidate Selection* module (where candidate indexes, MVs are selected based on a per query analysis), as well as the *Enumeration* module (where the search for the final configuration is performed over all candidates). These extensions are detailed in Section 6.

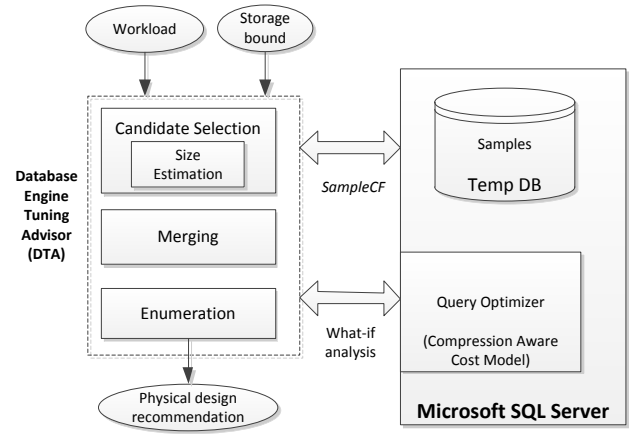

**Figure 1 Overview of Compression Aware Database Designer.**

In Section 7 we empirically evaluate our techniques on the TPC-H benchmark workload as well as a real world workload. We conclude and discuss future work in Section 8.

## 4. Index Size Estimation Methods

As described earlier, efficient estimation of the size of a compressed index is crucial to physical database design. This section explores efficient methods to estimate the compressed index size. We first extend the existing *SampleCF* method [11] (described in Section 2.2) to reduce the cost of sampling. Next, we propose new *deduction* methods (Section 4.2) that can infer the compressed index size based on sizes of other indexes whose sizes are already known. Finally (in Section 4.3), for *SampleCF* as well as the new deduction methods, we empirically quantify the distribution of errors in size estimation that we observe over a large variety of datasets and indexes.

### 4.1 Extending *SampleCF*

*SampleCF* performs size estimation based on random sampling. However, taking a uniform random sample from a large table is expensive. Since a physical design tool can consider a large number of indexes for a workload (e.g. thousands of indexes for complex workloads is common), taking a random sample for estimating the size of each index is infeasible. Therefore, we propose to amortize the sampling cost across all indexes on a given table by taking a random sample only once per table, and reusing it for all indexes on that table. For partial indexes and a certain class of materialized views (MVs) with foreign-key joins and grouping, we maintain special samples based on filtering and *join-synopses* [2], a sample of pre-joined tables. For more details about this, we refer readers to **Appendix B**.

We empirically observed that amortizing the sampling cost reduces the cost of *sampling* by a few orders of magnitudes. Consequently, now the cost of *creating an index* on the sample becomes a significant cost. We therefore develop index size estimation methods that can avoid invoking *SampleCF* altogether.

## 4.2 Deducing Index Size

In this section we present techniques for deducing the size of a compressed index based on *other* indexes whose sizes are known. The deduction technique incurs virtually no cost to estimate the size of an index. We describe three deduction techniques for different types of compression scheme.

**Types of Compression**: The way we deduce the index size depends on the type of compression scheme. We categorize the various compression schemes introduced in the background into two groups; *Order-Independent* (**ORD-IND**) and *Order-Dependent* (**ORD-DEP**). ORD-IND compressions such as NULL-suppression and global dictionary encoding have the same compressed size regardless of the order of tuples in the index page while ORD-DEP compressions such as local dictionary encoding and run length encoding (RLE) are sensitive to the order of tuples, or the value distribution in each page.

For example, suppose two columns A, B and compressed indexes on them $I^C_{AB}$, $I^C_{BA}$. As illustrated in Figure 2, the order of tuples in the two composite indexes is quite different. However, NULL-suppression suppresses the same total number of NULLs in both cases. Likewise, global dictionary encoding constructs the exactly same dictionary for the two indexes and replaces the same number of entries with pointers to the dictionary (assuming the DBMS constructs a dictionary *per column*).

**Column Set Deduction (ORD-IND)**: Thus, the first deduction method, as we call *Column Set Deduction* (*ColSet*), deduces the size of $I^C_{AB}$ from that of $I^C_{BA}$ as Size($I^C_{AB}$)=Size($I^C_{BA}$) because the order of data does not affect the compressed size. More generally, every two indexes compressed using a method in ORD-IND have the same size if they contain the same **set** of columns. *ColSet* deduction is particularly useful for clustered indexes. All clustered indexes on the table have the same compressed size because all of them contain the same set of columns. Hence, we can avoid *SampleCF* for all but one compressed clustered index per table.

**Column Extrapolation (ORD-IND)**: *Column Extrapolation* (*ColExt*) estimates the size of a composite index from subsets of the index. Suppose we want to estimate Size($I^C_{AB}$) and we know Size($I^C_A$) and Size($I^C_B$). Let $R(I_{AB})$ be the size reduction achieved by compressing $I_{AB}$, i.e., $R(I_{AB})$ = Size($I_{AB}$) - Size($I^C_{AB}$). If the compression is ORD-IND, we can estimate $R(I_{AB})$ from $R(I_A)$ and $R(I_B)$ as $R(I_{AB})=R(I_A) + R(I_B)$ because ORD-IND achieves the same size reduction for each column. Hence, Size($I^C_{AB}$)=Size($I_{AB}$) - $R(I_A)$ - $R(I_B)$.

**Column Extrapolation (ORD-DEP)**: It is also possible to use the idea of column extrapolation for ORD-DEP compression such as page-local dictionary encoding, but we cannot simply sum up the reduction in this case. As shown in the figure, the order of values of A in $I_{AB}$ is same as $I_A$ while that in $I_{BA}$ is *fragmented* by the leading column B, reducing the number of repeating values of A in each page.

To account for the fragmentation, we estimate the average number of distinct values in each page and penalize the size reduction attributed to following columns.

Let $DV(I_X, Y)$ be the average number of distinct values of column Y and $T(I_X)$ be the number of tuples in a page of index $I_X$. Then, the average fraction of Y replaced by the dictionary are defined as $F(I_X, Y) = (T(I_X) - DV(I_X, Y))/T(I_X)$. For example, $T(I_{AB})$=4,

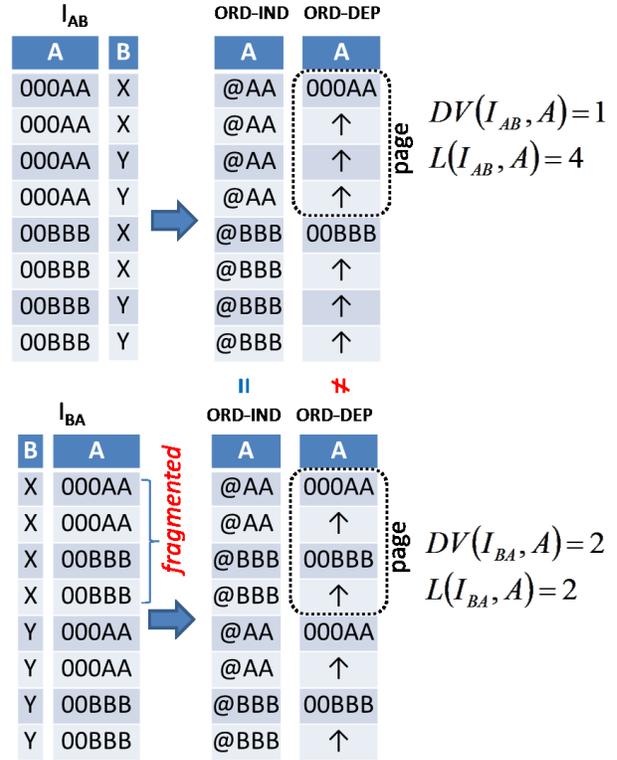

**Figure 2. Order Independent/Dependent Compression.**

$DV(I_{AB}, A)$=1, and $F(I_{AB}, A)$=3/4 of the values of A were eliminated.

Suppose we deduce the size of $I_{BA}$ from $I_A$ and $I_B$, so we know $R(I_A)$ and $R(I_B)$. As the space saving of compression is linear to the number of values replaced by the dictionary, $R(I_{AB}) = R(I_B)F(I_{BA}, B)/F(I_B, B) + R(I_A)F(I_{BA}, A)/F(I_A, A)$.

As B is the leading key of $I_{BA}$, its value distribution in pages is equal to that of $I_B$, thus $F(I_{BA}, B) = F(I_B, B)$. As for A, its value distribution is fragmented by B thus $F(I_{BA}, A) < F(I_A, A)$. To calculate $F(I_{BA}, A)$ and $F(I_A, A)$ (in other words $DV(I_{BA}, A)$ and $DV(I_A, A)$), we consider the average *run length* of a value of A in $I_{BA}$ and $I_A$. Let $L(I_X, Y)$ be the average run length of a value of Y in $I_X$. For example, $L(I_{BA}, A)$=2, $L(I_A, A)$=$L(I_{AB}, A)$=4 in the figure. We approximate the values with cardinality statistics as:

$$L(I_A, A) = \lfloor \#TotalTuples/|A| \rfloor, L(I_{BA}, A) = \lfloor L(I_A, A)|A|/|AB| \rfloor$$

The approximated values are actually $L(I_A, A)$=8/2=4 and $L(I_{BA}, A)$=4*2/4=2. Note that, in order to calculate $L(I_{BA}, A)$, we do not simply divide $L(I_A, A)$ by $|B|$ because A and B might be *correlated*, i.e., $|A|/|AB| << |B|$.

Then, we approximate the number of distinct values as follows. If $L(I_X, Y)>1$, $DV(I_X, Y) = \lceil T(I_X)/L(I_X, Y) \rceil$ (e.g., $DV(I_{BA}, A)$=4/2=2). Otherwise, $DV(I_X, Y) = |Y| - |Y|pow(1-1/|Y|, T(I_X))$ which is the expected number of distinct sides when throwing a $|Y|$-sided dice $T(I_X)$ times.

In principle, this estimation is also applicable to RLE compression although we have not empirically evaluated it for RLE.

## 4.3 Accuracy of Estimation Methods

Deduction effectively enables us to eliminate some *SampleCF* calls and thus reduces the cost of index size estimation. However both *SampleCF* and deduction can result in size estimation errors.

To analyze the errors of size estimation, we empirically evaluated *SampleCF* and deduction against hundreds of indexes on various datasets and skew-ness (details in **Appendix C**). In summary, we observed consistent behaviors across all datasets that we tried. For *SampleCF*, as expected, we observe that the average and variance of errors are higher with smaller sample size. Also, deductions introduce more errors when we extrapolate more indexes. This analysis of errors in compressed index size estimation provides a basis for the optimization framework described in next section.

## 5. Optimizing Index Size Estimation

A physical database design tool may need to compute sizes of a large number of compressed indexes. Inefficient size estimation can make the runtime of the tool unacceptable. In fact, we empirically observed that index size estimation without exploiting the *deduction* methods (Section 4.2) causes a dominating runtime overhead on a database design tool (see **Appendix D** for experiments). Thus, given a large set of compressed indexes whose sizes need to be estimated, we need to find a good strategy. Such a strategy can consist of using *SampleCF* for some indexes (more expensive but more accurate) and using *deduction* methods (much faster but less accurate) for others. Since we want size estimation to have low error, we need to balance this trade-off between accuracy and performance. In this section, we formulate the problem as an optimization problem and devise a graph search algorithm to solve it.

### 5.1 Problem Statement

The problem of index size estimation is defined as follows.

> **Inputs**: A set of compressed indexes whose sizes need to be estimated (*targets*), a tolerable error ratio $e$ and a confidence parameter $q$ such that the estimated sizes of the targets have errors less than $e$ for at least $q$ probability.
> **Output**: Sampling ratio $f$ (fraction of table to sample) and the size estimation method to use for each index (*SampleCF* or *deduction*) such that the total cost of size estimation is minimized without violating the accuracy constraint.

For example, when $e$=20% and $q$=95%, the estimated size of a compressed index whose true size is 100 MB must be between 120MB and 83.3MB for at least 95% probability. Higher $e$ and lower $q$ will allow a smaller sample size and more deductions, therefore is faster at the cost of accuracy. In order to determine whether an estimate satisfies the accuracy requirement, we quantify its error as follows.

**Bias and Variance of Error**: Every sample-based size estimation approach can have a potentially arbitrary error for a particular index. However, we can analytically infer or empirically compute the *expected* error (bias) and its variance. For example, prior work showed that *SampleCF* for NULL suppression encoding is unbiased and has at most $1/rf^2$ variance where $f$ is sampling ratio and $r$ is the number of sampled tuples [11]. We devised similar formulas for all compression types and deduction methods based on empirical analysis (for more details, see **Appendix C**).

**Composition of Errors**: Let $X_A$ be the random variable to denote the result of size estimation for $I_A$ divided by its true size, thus $X_A$=1 is the most accurate estimation. Suppose we deduce the size of $I_{AB}$ from $I_A$ and $I_B$ with *ColExt*. To account for amplified errors by deduction, we formulate the deduced result as $X_{AB} = X_A X_B X_{ColExt}$ where $X_{ColExt}$ is the random variable to denote the result of the deduction for perfectly accurate inputs (sizes of $I_A$ and $I_B$). The variance of such a product of random variables is calculated as $\prod_i \left(V(X_i) + E(X_i)^2\right) - \prod_i \left(E(X_i)^2\right)$ [9] while the expected value is simply the product of each expected value assuming independence among the random variables. We note that the above formula is only a heuristic if $X_i$s are not truly independent (e.g. that can happen if we reuse the same sample for computing sizes of $I^C_A$ and $I^C_B$). Then, we define the probability that the error of the estimation is within $e$ as the integral of normal probability distribution between $[1/(1+e), 1+e]$ with the bias and variance. We assumed normal distributions based on our empirical analysis, but any parametric distributions can be used instead.

**Size Estimation Cost**: We model the cost of index size estimation as the amount of data we need to index. The cost of *SampleCF* on an index is considered as the number of data pages in the index before compression. Hence, *SampleCF* on wider indexes with larger samples costs more. The cost of deduction is zero.

**Existing Indexes**: The database might already have a compressed index before running the database design tool. Such an index provides a perfectly accurate size of itself simply from the database statistics. Hence, we consider that such an index has zero bias and variance as well as zero cost for size estimation.

### 5.2 Graph Search Algorithm

We solve the problem as a directed graph problem illustrated in Figure 3. The graph has two types of nodes; index nodes and deduction nodes. An index node (e.g., "AB") denotes the size estimation for an index and has one of three states; *NONE*, *DEDUCED* and *SAMPLED*. NONE is the initial state of all index nodes where we have not yet made a decision for that index. DEDUCED and SAMPLED mean we estimate the size by *deduction* and *SampleCF* respectively. Edges connect index nodes from/to deduction nodes. We call the node from which an edge is coming as a *child* node and the node at which the edge is directed as a *parent* node.

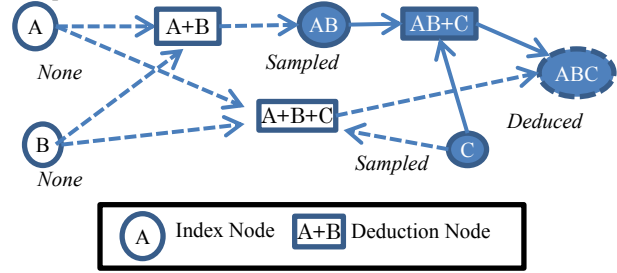

**Figure 3. Graph of Index and Deduction Nodes.**

A deduction node represents a possible deduction to estimate the size of its parent based on its children. For example, the deduction node "A+B" has a parent index node "AB" (the index whose size can be deduced) and child index nodes "A" and "B" (indexes using which deduction can be performed). A deduction node is enabled only when all its children are DEDUCED or SAMPLED, i.e. their sizes are known.

The goal is to find an assignment of the states to each node such that all target indexes are marked as DEDUCED or SAMPLED and also satisfy the desired accuracy i.e., error constraints.

Suppose $I_{ABC}$ and $I_{AB}$ are the target indexes. The solution in the figure is to *SampleCF* on $I_{AB}$ and $I_C$ and then deduce the size of $I_{ABC}$ from them. Compared to *SampleCF* on $I_{AB}$ and $I_{ABC}$, this solution gives less accuracy on the size estimation of $I_{ABC}$ because it is deduced. However, because building a sample composite index on ABC costs more than on C, the solution is better unless the error constraint is too tight to allow the deduction. Another possible solution is to *SampleCF* on all singleton indexes and deduce the size of $I_{AB}$ and $I_{ABC}$. In that case, there are two options to deduce the size of $I_{ABC}$; A+B+C and AB+C.

An exact algorithm to get the optimal solution takes time exponential in the number of indexes. Instead, we developed a greedy heuristic algorithm shown below which achieves a high quality and yet is much faster. We start from narrow indexes and greedily determine the state of the index (Line 3), deducing the size from already determined narrower indexes if possible (Line 6-7). Otherwise, we sample the index (Line 11) unless changing only a few of the narrower indexes from DEDUCED to SAMPLED satisfies the accuracy constraint (Line 8-9). For each target index, this algorithm only considers changing the state of the index and its direct children, thus it finishes very quickly even for a large number of indexes.

Finally, for choosing a suitable sampling fraction $f$, we try several different values of $f$ and pick the $f$ for which the greedy algorithm produces a solution with the smallest total cost. Note that certain combinations of $f$, $e$ and $q$ can give an invalid result, e.g. even applying *SampleCF* on all targets does not satisfy the accuracy constraint. As demonstrated in the experimental section, this simple algorithm achieves sometimes orders of magnitude smaller total cost while maintaining a good accuracy of size estimation.

**Greedy Algorithm**
1. Add existing indexes to the graph with SAMPLED state.
2. Add target indexes to the graph with NONE state;
3. foreach(target) { // from narrower to wider
4.   Add all child deduction nodes of this node to the graph;
5.   Add children of the deduction nodes, if not yet added;
6.   if (any child deduction satisfies the constraint with the given $f$, $e$ and $q$) {
7.     Mark this node DEDUCED from the deduction node; (if multiple deductions are eligible, pick the one with the highest probability)
8.   } else if (any deduction can be enabled by doing SampleCF on its children such that the sum of their costs is lower than the cost of sampling this node) {
9.     Mark this node DEDUCED from the deduction node and mark its children SAMPLED; (if multiple deductions are eligible, pick the one with the least cost)
10.   } else {
11.     Mark this node SAMPLED;
12. }}
13. foreach (enabled index) //from wider to narrower
14.   if (not targeted nor used by parents)  Remove the node;

## 6. Handling Space-Time Tradeoff

As discussed earlier, compressed indexes can greatly amplify the space-time tradeoff that automated physical design tools need to consider. Thus, the quality of physical design solutions produced by these tools can potentially improve by leveraging new techniques for handling this tradeoff. For instance, Microsoft SQL Server's design tool (DTA) first separately analyzes each query in the workload and from the space of all syntactically relevant indexes for the query, it selects a set of *candidate* configurations (the *Candidate Selection* step). The final configuration is then picked from the union of candidate configurations over all queries in the workload (the *Enumeration* step). This is illustrated in Figure 4. However, we found that such an approach can miss good physical database designs that fully exploit the benefits of compressed indexes (discussed below in Sections 6.1 and 6.2). Thus revisiting the pruning heuristics in these tools can become important for compressed indexes. Although our solutions in this section are described in the context of a specific physical design tool (DTA), the key ideas are also applicable to other design tools.

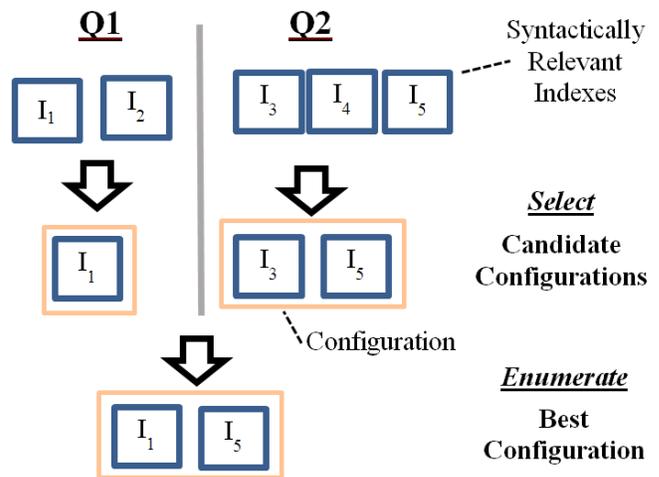

Figure 4. Candidate Selection and Enumeration steps in DTA.

### 6.1 Candidate Selection

The number of syntactically relevant indexes for a query can be quite large even though few of them are actually useful. Hence, a design tool usually selects a few small candidate configurations by picking the top-k configurations (e.g. k=2) that with the lowest optimizer estimated cost for each query. This best-per-query approach works well with a large space budget, but in a tight space budget it could result in designs that speed up only a small number of queries. This is because the approach might not capture space efficient indexes that are not the best in terms of query cost, which might achieve lower overall cost for the workload since they allow more indexes to be selected for other queries. Compression makes this space-performance trade-off even more prominent. Compressed indexes are often not the best indexes for a query because of their decompression CPU costs. Thus the current approach can miss out many useful compressed indexes except indexes that compress sufficiently to overcome the decompression cost with the reduced I/O cost.

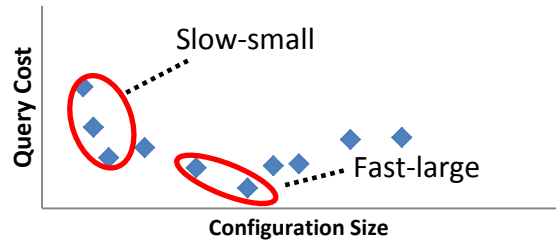

Figure 5. Skyline Candidate Selection.

We therefore developed the *Skyline* method for candidate selection. Rather than choosing only the top-k configurations for a query with lowest cost, we pick all configurations in the skyline of size and query cost. The idea is to capture a spectrum of indexes ranging from fast-large to slow-small as illustrated in Figure 5. To construct the skyline for each query, we compute the cost of all candidate configurations considered by the tool. Then, for each of them, we test if there is another configuration that dominates it, i.e. has lower cost and is also smaller. If so, we remove the configuration from the skyline. The overhead to construct the skyline is $O(n^2)$ where $n$ is the number of configurations for each query. We observed that the overhead is negligible compared to obtain the optimizer estimated cost for these $n$ configurations. In the experimental section, we demonstrate that the skyline selection along with the backtracking described in next section significantly improves quality of physical design especially for tight space budgets.

Although the skyline selection improves the design quality, it produces more configurations which cause more computation in the enumeration phase (Section 6.2). As a compromise between design quality and design time, one possible extension for large complicated query workloads is to pick a small number, not all, of configurations among the skylines by clustering them into groups and selecting a representative configuration from each group.

## 6.2 Enumeration

Across all indexes from all candidate configurations, the goal of enumeration is to choose the best set of indexes that speed up the entire query workload and also fit the space budget. Since there are an exponential number of combinations of indexes, it is infeasible to search for the exact optimal set. Hence, most design tool employs a greedy approach (e.g. [7] [15]) which picks the next index that reduces the cost the most at each step, starting from an initial configuration. Although this pure greedy approach is fast and scalable, we found realistic cases involving compressed indexes where this approach can result in poor solutions. Consider the example in Figure 6.

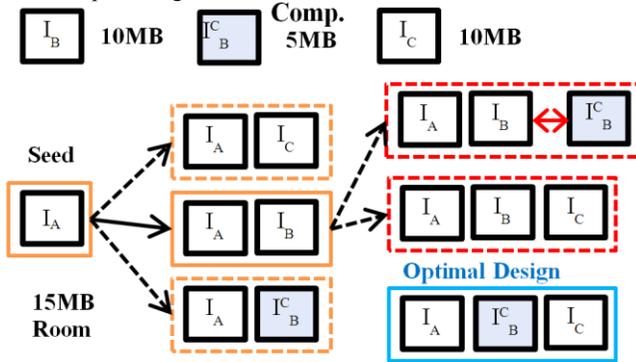

Figure 6. Greedy Algorithm with Compressed Indexes.

The greedy algorithm adds the index that reduces the workload cost the most at each step. In Figure 6, at the first step adding $I_B$ turns out to be the best option. However, at the next step, we have only 15-10=5MB of remaining space budget. Adding $I_C$ will be oversized, but adding the compressed index $I^C_B$ is not useful because we already have the faster $I_B$ without compression. Thus, although the best design is actually $I^C_B$ and $I_C$, the greedy algorithm never reaches the solution. The above situation can often occur with indexes that compress heavily such as clustered indexes because they may save a lot of space but may also perform slowly with queries. Since a table can have only one clustered index, the pure greedy approach cannot improve the design once an uncompressed clustered index is chosen.

A similar problem is caused by *competing* indexes which speed up the same queries but only one of them can be used at the same time just like $I_B$ and $I^C_B$ in the above example. Some design tools e.g. [15] consider the *density* at each greedy step, i.e. choosing the index that has the highest ratio of "benefit" to size. Figure 7 illustrates how it works. For simplicity, suppose there is only one query. Assume $I_B$, $I^C_B$ and $I_C$ speed it up for 10, 8 and 5 seconds respectively. The density of them at the first greedy step is 10/10=1, 8/5=1.6 and 5/10=0.5. Thus, $I^C_B$ is picked at this step. At the next step, the benefit of adding $I_C$ is still 5 seconds while that of adding $I_B$ is only 2 (=10-8) seconds because we already contain the slower but competing index $I^C_B$. The density of $I_B$ and $I_C$ are 2/10=0.2 and 0.5, thus $I_C$ is picked at this step, resulting in the optimal design.

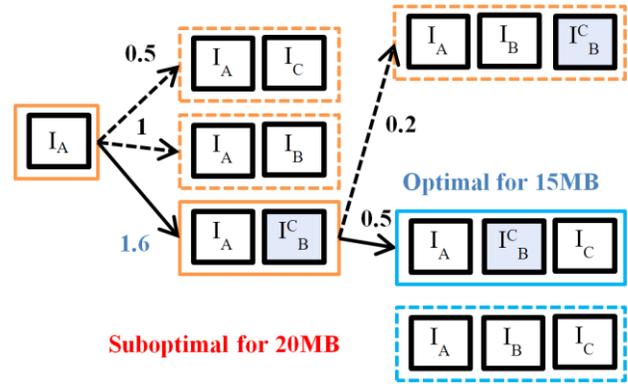

Figure 7. Density-Based Greedy with Compressed Indexes.

However, the density based greedy results in the same design even for 20MB space budget where the optimal design is $I_B$ and $I_C$. Also, we find that a density based approach tends to add many small but not so beneficial indexes which often cause a suboptimal design for larger budgets.

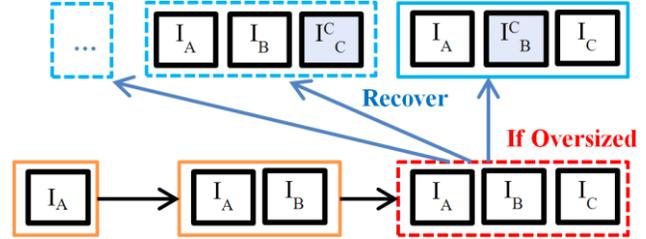

Figure 8. Backtrack to Recover an Oversized Greedy Choice.

In order to capture a good design in both tight and plenty space budgets, we add a backtracking phase to the pure greedy approach illustrated in Figure 8. It works just like the pure greedy until a greedy choice exceeds the space budget. Such an oversized configuration was not considered in the original greedy, but we try to recover it by replacing one or more indexes in the configuration with its compressed variant. We consider replacing each index and choose the replacement that performs fastest while making the configuration below the budget. Then, we compare the recovered configuration with other greedy choices as usual.

Finally, we note that some physical design tools *merge* indexes to generate candidate objects that benefit more than one query [8] (see also Figure 1). Our design tool generates compressed variants of such merged objects too, but we have not yet carefully studied how compression could affect merging, e.g., adding or removing some columns from the merged object might improve the compression fraction. Revisiting the problem of index merging in the context of compression could have significant impact on quality of database design as well.

## 7. Experiments

We now present empirical analyses on performance and quality of our compression aware design tool. Due to limited space, this section only provides a summary of the findings. We refer readers to **Appendix D** for the full details of our experiments.

We have implemented our techniques on Microsoft SQL Server 2008 R2, modifying its query cost models to account for compression and decompression CPU costs (for more details, see **Appendix A**). We also modified the SQL Server's Database Tuning Advisor (**DTA**); we refer to our compression aware physical database design tool as (**DTAc**). We run DTAc and DTA and evaluate them for two workloads: TPC-H and a real world customer database (*Sales*) which track sales of a particular company. In both workloads, we also vary the *weights* of the bulk load statements to represent SELECT intensive workloads and INSERT intensive workloads. Simply put, a database design with more indexes and heavier compression is suited for SELECT intensive workloads while a database design with less indexes and lighter compression is suited for INSERT intensive workloads because of the overheads to maintain indexes against INSERTs.

## 7.1 Results

**Size Estimation for Compressed Indexes**: We first evaluated the index size estimation framework (Section 5) alone against target indexes considered in TPC-H. With a tight accuracy requirement, the deduction strategy suggested by the estimation framework achieves 3 to 10 times smaller estimation cost than applying *SampleCF* on every index. With a looser accuracy requirement, the speed up becomes as large as 50 times because our framework can aggressively use deductions (Section 4.2). We observed that the strategy costs on average only 8% more than the optimal strategy obtained by an exact algorithm. Our greedy algorithm finishes within a second for more than 300 indexes while the exact algorithm does not finish in hours.

Next, we compare the running time of DTAc with and without the deductions. We observe that deductions actually reduce the overhead of index size estimation from dominating to not significant compared to the runtime of the original DTA. The real speed up of the size estimation overhead is a factor of 3.

We observe that the actual accuracy of index size estimation have less than 10% error in most cases. These results show that our size estimation module accurately and efficiently estimates the size of compressed indexes by automatically choosing the best sampling ratio and deduction strategy for the given user requirements.

**Candidate Selection and Enumeration**: Second, we verify the effects of the new candidate selection and enumeration techniques for compressed indexes. We run DTAc turning on/off the Skyline selection and Backtracking in enumeration. We find that, although all versions of DTAc generate compressed variants of indexes as candidates, only DTAc with both Skyline and Backtracking achieves significantly better designs especially in tight space budgets (up to a factor of 2). This is because the current candidate selection which picks only a few best configurations per query cannot capture the potential of compressed indexes with smaller sizes; and the current enumeration algorithms cannot choose an index that is slower but saves space.

**Comparison with no compression**: Then, we compare designs produced by the full implementation of DTAc with the DTA on TPC-H and the Sales database. In most cases, designs produced by DTAc are faster for a factor of 1.5 to 2 because DTAc utilizes compression to make indexes faster and also to allow more indexes within the space budget. The difference is smaller in larger space budgets (10%-50%) because more indexes can fit the space budget without compression. Also, in the INSERT intensive cases, DTAc appropriately avoided compressing too many indexes, being aware of the overheads of compressed indexes. This prevents the generated design from slowing down with larger budgets, which we actually observed with a naïve design tool that decouples compression from the choice of indexes.

## 8. Conclusion and Future Work

Data compression in DBMS has a potential to reduce both space consumption and I/O costs at the expense of CPU overhead for compression and decompression. The trade-offs of compression make the job of physical database design even harder. In this paper, we identified technical challenges in considering compressed indexes in a database design tool and developed techniques to address these challenges. We implemented our techniques inside a commercial DBMS engine and its physical design tool. Our empirical results suggest that the modified design tool achieves significantly better design quality compared to the unmodified design tool without adding too much overhead.

One open problem is physical design for Column-Store which utilizes compression more heavily and flexibly [1]. For example, RLE can make column data several orders of magnitude smaller and thus faster to read, but it is quite sensitive to the sort orders. Developing a design tool that fully exploits the potential of compression in Column-Store is interesting future work.

# Appendix
## A. Compression-Aware Cost Model
An index (or an MV) affects the performance of the database either positively or negatively. Typically, it speeds up *reads* (SELECT) while it slows down *updates* (INSERT/DELETE/UPDATE). The standard approach in automatic database design, called *What-If* analysis [7], is to analytically quantify the benefits of having each candidate index by calling the database's query cost models and choose a set of indexes that achieve the largest benefits overall.

Therefore, in order to let the database design tool consider the effects of compressing indexes, we need to modify the query cost models of the database for both reads and updates.

In this appendix section, we describe the way Microsoft SQL Server compresses and decompresses data on indexes and explain how we model the CPU overheads of the operations. Although we did not have a chance to take a look at internals of other commercial databases, we believe the cost models are applicable to them too because their compression scheme and basic mechanisms to handle compressed data are similar to ours.

### A.1) Cost Model for Updates
SQL Server compresses data when some update operation (e.g., INSERT) modifies a page. SQL Server has two packages of compressions; ROW (null-suppression) and PAGE (local dictionary and prefix encoding). ROW is an ORD-IND compression while PAGE is an ORD-DEP compression. As PAGE has higher overheads to compress, SQL Server delays applying PAGE compression even if the page belongs to a PAGE compressed index. Such a page is first compressed with ROW compression, and then again compressed with PAGE compression when the page is "done" with modifications (becomes full or ejected from the bufferpool).

We adjust the cost model for update operations on compressed indexes in SQL Server as follows.

$$\text{CPUCost}_{update} = \text{BaseCPUCost} + \alpha * \#\text{tuples}_{written}$$

where BaseCost is the existing cost model for the update operation and $\alpha$ is a constant defined for each compression type which represents the CPU cost to compress the tuple (larger for PAGE compression). We determine the value of $\alpha$ based on the micro benchmark in [13].

### A.2) Cost Model for Reads
When reading data in compressed indexes, SQL Server retrieves the index pages from the disk and keeps them compressed in the bufferpool to save memory consumption, decompressing the buffered page each time the page is read. Therefore, a read operation on a compressed index causes the same CPU overhead for decompression no matter how many pages of the index reside in the bufferpool.

However, SQL Server avoids decompressing unused columns in the index page. It decompresses only the columns that are projected, predicated or aggregated by the query. Let $\#\text{columns}_{read}$ be the number of such used columns in the query. The cost model for read operations on compressed indexes is defined as follows.

$$\text{CPUCost}_{read} = \text{BaseCPUCost} + \beta * \#\text{tuples}_{read} * \#\text{columns}_{read}$$

Where $\beta$ is a constant that represents a cost of decompressing one column data of one tuple (again, higher for PAGE compression) which is determined by benchmarking.

We note that our model of I/O cost is unchanged, but the smaller (estimated) size of compressed indexes implicitly handles it.

## B. Samples for Partial Indexes and MVs
In this appendix section, we describe extensions to our size estimation module for partial indexes and materialized views.

### B.1) Filtered Samples
As described in Section 4.1, our size estimation framework maintains sample tables to apply *SampleCF* on. Although the *base* sample tables are sufficient for *SampleCF* on simple indexes, they do not work for more complex indexes that contain WHERE clauses (partial indexes), JOINs and/or GROUP-BYs (indexes on MVs). For this reason, our framework also maintains *filtered samples* and *MV samples*.

A filtered sample is generated by applying the WHERE clause on the base sample table and used for partial indexes. For example, suppose a partial index "**CREATE INDEX I$_1$ ON LINEITEM (SuppKey) WHERE SuppKey<2000**". We run the following SQL to construct a filtered sample for it.
  SELECT * INTO S$_{I1}$ FROM S$_{LINEITEM}$ WHERE SuppKey<2000
where S$_{LINEITEM}$ is the sample table of LINEITEM. This filtered sample gives an accurate estimation as far as S$_{LINEITEM}$ is uniformly random (not skewed with respect to the WHERE clause) and contains a reasonably large number of tuples.

### B.2) Join Synopses
An MV sample, on the other hand, is more difficult to construct for two reasons. The first difficulty is JOIN. Suppose the following MV which joins LINEITEM with SUPPLIER.
  CREATE VIEW MV1 AS SELECT SuppKey, Price, SuppCity FROM LINEITEM JOIN SUPPLIER ON (SuppKey)
A naïve way to take a sample for this MV is to join two sample tables as follows.
  SELECT SuppKey, Price, SuppCity INTO S$_{MV1}$ FROM S$_{LINEITEM}$ JOIN S$_{SUPPLIER}$ ON (SuppKey)
However, this usually results in very few tuples in the MV sample because each base sample is randomly taken and might not have tuples that match the foreign key values. To address this problem, we construct *join synopses* [2] of the database, which is applicable for Key-Foreign Key join views.

When the framework is initialized, it takes a random sample of fact tables (e.g., LINEITEM). Next, it joins the sample fact table with the *original* dimension tables so that foreign key values have always matching tuples. The result is a very wide joined sample. We use such join synopses to create MV samples when the database design tool requests them. For instance, we take an MV sample for MV1 by running the same SQL above but on the joined synopses. Then, we construct compressed indexes on the sample to estimate the compressed size of indexes on the MV.

### B.3) MVs with Aggregation
Another important case is materialized views with GROUP BY and aggregation. To estimate the size of a compressed index, we also need to know the number of entries (tuples) in the index. Although we can simply use the base table's statistics for simple indexes, we need to estimate how many distinct groups the MV will have. Suppose the following MV and its MV sample.
  CREATE VIEW MV2 AS SELECT ShipDate, SUM(Price) FROM LINEITEM GROUP BY ShipDate
  SELECT ShipDate, SUM(Price) INTO S$_{MV2}$ FROM S$_{LINEITEM}$ GROUP BY ShipDate
Here, S$_{MV2}$ has about 1,000 tuples. If the number of tuples simply scales up to the sampling ratio (S$_{LINEITEM}$ contains 1% of LINEITEM), the MV would have about 100K tuples. However, the actual number of tuples in the MV is only 2,000; the number of distinct SHIPDATE values. This example illustrates, unlike partial indexes, we need to consider the distribution of distinct values to estimate the number of tuples in MVs.

The obvious way to get the correct answer is to run a query "**SELECT COUNT (DISTINCT ShipDate) FROM LINEITEM**", but running such a query for every candidate MV in the database design tool is prohibitively expensive. Another way is to ask the query optimizer to estimate the number of tuples returned by the query that defines the MV. Query optimizer answers the estimate based on statistics of each column. However, this estimate is often inaccurate because MVs usually aggregate on more than one column and the optimizer simply assumes independence between the columns unless we additionally scan the table and collect multi-column statistics.

**Algorithm CreateMVSample ()**
1. *SELECT <MV-Project>, COUNT(*) AS cnt INTO $S_{MV}$ FROM <join-synopses> WHERE <MV-WHERE> GROUP BY <MV-GROUP BY>.*
2. *r = SELECT SUM(cnt) FROM $S_{MV}$*
3. *d = SELECT COUNT(*) FROM $S_{MV}$*
4. *FilterFactor = r / <join-synopses>.#tuple*
5. *n = RootTable.#tuple * FilterFactor*
6. *f = SELECT cnt AS frequency, COUNT(*) AS value FROM $S_{MV}$ GROUP BY cnt*
7. *MV.#tuple = **AdaptiveEstimator**(f, d, r, n);*

We devised a new algorithm shown above to address this issue without adding overheads to the design tool. Typically, DBMS requires an MV with aggregation to always contain a COUNT(*) column in its definition (or internally add as a hidden column) for incremental maintenance. DBMS increases or decreases the counter when a newly inserted or deleted tuple falls into the group and eliminates the group when the counter gets to zero. We utilize this information as *frequency statistics* for distinct value estimators.

A distinct value estimator, for example *Adaptive Estimator* [6], gives an estimated number of distinct values based on frequency statistics $f = \{f_1, f_2, \ldots f_k\}$ where $f_k$ is the number of distinct values that appear $k$ times in the random sample. We get the statistics by querying on the MV sample and aggregating on the COUNT column. We additionally compute *r* and *d*, the number of tuples in the MV sample **before** and **after** the aggregation respectively as well as *n*, the number of tuples in the original table. The Adaptive Estimator, which is implemented in our database design tool, takes these as inputs and gives the estimated number of tuples in the MV. We keep these estimates for each MV sample we took.

**Table 1. Average Errors of #Tuples in Aggregated MVs.**

| Optimizer | Multiply | AE |
|---|---|---|
| 96% | 379% | 6% |

Table 1 compares the average errors of the three methods to estimate the number of tuples of all MVs with aggregation considered by DTA for TPC-H.

*Optimizer* is to ask the query optimizer to estimate the cardinality of the MV based on single-column statistics. *Multiply* is to simply multiply the number of distinct values in random samples with sampling ratio. As expected, both of the two methods have large errors. The optimizer estimate is better, but still the error is 96% (error of a factor of 2) on average. Unlike the others, our algorithm using Adaptive Estimator (*AE*) achieves as low as 6% errors on average. This result demonstrates that the algorithm gives orders of magnitude more accurate estimates for the size of MV indexes. We also observe that its overhead is negligible.

## B.4) Indexes on Join Synopses
Additionally, we build indexes on the join synopses to speed up querying on them for creating MV samples. Although the sample tables are only the part of the original tables (e.g., 1%), the design tool has to apply joins and filters on them for each MV candidate. We found that indexes on primary keys and foreign keys significantly speed up this process.

## C. Analysis on Estimation Error
In this appendix section, we provide a detailed analysis on the accuracy of the index size estimation methods and their stochastic formulation used in our size estimation framework.

To quantify the errors of *SampleCF*, we applied *SampleCF* on hundreds of indexes considered for TPC-H. Figure 9 plots the average bias and standard deviation of local dictionary (LD) and NULL-suppression (NS) for a few *f*. Both bias and standard deviation drop very quickly as *f* increases, except bias of NS which is always very low as expected in [11]. We formulated the errors of *SampleCF* by applying the least square error estimation on this data with an assumption that bias and standard deviation becomes zero when f=1 (full index creation). We repeated the same analysis on the skewed version of TPC-H and the TPC-DS benchmark to see the stability of our formulation. Table 2 shows that the parameters of the error formula are quite stable between different table scheme and data skews. We also analyzed the shape of error distributions in each dataset and observed that they are close to normal distributions.

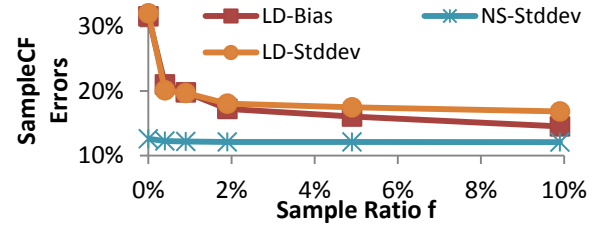

**Figure 9. Error Bias and Variance of SampleCF.**

**Table 2. Least Square Error Analysis on Various Data Sets.**

| SampleCF | **LD-Bias** | **NS-Stddev** | **LD-Stddev** |
|---|---|---|---|
| TPC-H Z=0 | -0.015 ln(*f*) | -0.0062 ln(*f*) | -0.018 ln(*f*) |
| TPC-H Z=1 | -0.018 ln(*f*) | -0.0060 ln(*f*) | -0.017 ln(*f*) |
| TPC-H Z=3 | -0.013 ln(*f*) | -0.0056 ln(*f*) | -0.014 ln(*f*) |
| TPC-DS | -0.015 ln(*f*) | -0.0064 ln(*f*) | -0.017 ln(*f*) |

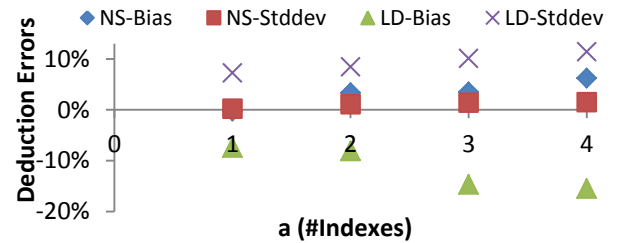

**Figure 10. Error Bias and Variance of Deduction.**

Similarly, Figure 10 shows the average bias and standard deviation of estimation errors we observed with column extrapolation deduction, plotted against the number of indexes from which to extrapolate (*a*); for example if size of $I^C_{AB}$ is extrapolated from $I^C_A$ and $I^C_B$, then the number of indexes from which to extrapolate is 2. Bias and standard deviation linearly grow with *a*. By fitting a line from the origin (no error with zero index to add), we formulated the error as shown in Table 3. Column set deduction always has a very low error. So, we assume it is unbiased and stable.

Note that our index size estimation framework (Section 5) can work for other compression and estimation methods if their errors can be characterized by parametric distributions with a given bias and variance.

**Table 3. Error Formula for Deduction.**

|  | Bias | Stddev |
|---|---|---|
| ColSet(NS) | 0 | 0.0003 |
| ColExt(NS) | 0.01 $a$ | 0.002 $a$ |
| ColExt(LD) | - 0.03 $a$ | 0.01 $a$ |

## D. Full Experimental Results

### D.1) Experimental Environment

We have implemented every technique we presented in this paper on Microsoft SQL Server 2008 R2. We have modified the query cost models in SQL Server to account for compression and decompression CPU costs as described in Section A and developed the modified physical database design tool (**DTAc**) based on the SQL Server's Database Tuning Advisor (**DTA**).

We run DTAc and DTA to compare their total query runtimes estimated by the SQL Server's cost models. All the experiments are done on a server running Windows 7 with Dual-core CPU, 4GB RAM and 10K RPM HDD.

### D.2) Datasets and Query workloads

We use two datasets and query workloads. The first is TPC-H Scale 1 which has 22 analytic queries and two bulk load INSERTs. Another is a real sales database (*Sales*) which has 50 analytic queries and two bulk load statements on fact tables. We vary the space budget from 10% to 100% of the database sizes without any indexes. It is interesting to note that DTAc might produce indexes even with 0% space budget by compressing existing tables (heaps/clustered indexes) and spending the saved space to secondary indexes.

In both workloads, we also vary the *weights* of the bulk load statements to represent SELECT intensive and INSERT intensive workloads. Simply put, a database design with more indexes and heavier compression is suited for SELECT intensive workloads while a database design with less indexes and lighter compression is suited for INSERT intensive workloads because of the overheads to maintain indexes against INSERTs.

### D.3) Results

**Size Estimation for Compressed Indexes**: We first verify the quality and runtime of the size estimation algorithm described in Section 4. Table 4 compares the quality (total sampling cost of suggested deduction strategy) of our greedy algorithm to 2 other methods with e=50%, q=90%. *All* simply applies SampleCF for all nodes. *Optimal* is an exact algorithm with recursion given below. As *Optimal* takes too long time for many indexes, we only used indexes in LINEITEM table in TPC-H. Also, we limited the number of columns per index to 7. As the result shows, *Greedy* achieves 2x to 6x smaller sampling cost compared to *All* by utilizing deductions. With a larger error tolerance such as e=100%, the difference becomes as large as 50 times. The quality of *Greedy* closely follows *Optimal*, costing on average only 8% and at most 30% more. Finally, *Greedy* never violates the accuracy constraint unless even *All* does. To contrast with this, we tested yet another simple algorithm which samples only singleton indexes and deduces all target indexes from them. Such an algorithm achieves the minimal sampling cost, but yields high errors (e.g., >4x mis-estimation for >50% probability, >2x mis-estimation for >90% probability).

As for runtime, *Optimal* did not finish in hours for all 300 indexes considered by DTAc while *Greedy* finished in a second.

**Optimal Graph Search Algo. (exact, but exponential)**
1. Construct subgraphs (clusters) of indexes which are considered together (e.g., AB and ABC and ABD);
2. foreach (*subgraph*) {
3.   *best* = NULL;
4.   Add all possible descendants and deductions;
5.   [Recursion] while *subgraph* is not empty {
6.    when it becomes empty, update *best* if it constitutes a satisfying assignment with least sampling cost so far;
7.    *branch* = widest remaining index in *subgraph*;
8.    Mark *branch* as SAMPLED, eliminate descendants that are no longer needed, then recurse;
9.    foreach (deduction to deduce the size of *branch*) {
10.     Mark *branch* as DEDUCED from the deduction, eliminate descendants no longer needed, then recurse;
11.   }}
12. Apply the assignments of *best* to the original graph.}

**Table 4. Quality (Cost) of Graph Algorithms. e=0.5, q=0.9.**

|  | f=1% | f=2.5% | f=5% | f=7.5% | f=10% |
|---|---|---|---|---|---|
| **All** | 222 | 555 | 1111 | 1666 | 2221 |
| **Greedy** | 114 | 284 | 393 | 589 | 352 |
| **Optimal** | 114 | 284 | 296 | 444 | 299 |

We then verify the real overheads of the size estimation in DTAc. Figure **11** compares the total runtime of DTAc on TPC-H with all features (clustered/secondary indexes, partial and MV indexes) with and without deduction. Note that even "w/o deduction" uses the Sample Manager described in Section 4 without which DTAc takes too long time to finish.

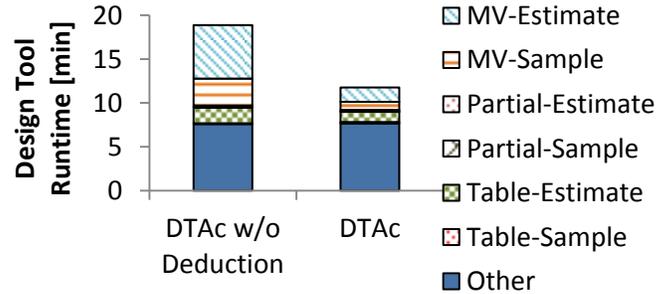

**Figure** 11 **Real Runtime of Index Size Estimation.**

All costs other than size estimation (e.g., optimizer calls, candidate generation and enumeration), denoted as "Other", are the inherent costs and largely same as the runtime of the original DTA which does not consider compressed indexes. The time to estimate the sizes of compressed indexes, denoted as "X-Estimate" (plain *Table* indexes, *Partial* indexes and *MV* indexes), significantly drops by utilizing deductions.

Overall, deductions reduce the overhead of index size estimation from dominating (700 seconds) to modest (200 seconds) compared to the rest of the tuning costs (500 seconds). For larger datasets (e.g., TPC-H Scale 10), the overhead of size estimation becomes even more dominant. We observed a similar result on *Sales*.

Regarding the actual accuracy of index size estimation, we observe that most cases have less than 10% errors. These results show that our size estimation module strikes a reasonable balance between accuracy and efficiency in size estimation of compressed indexes.

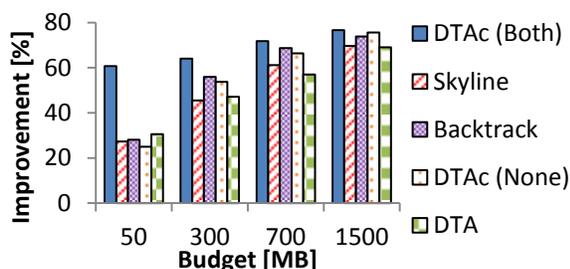

**Figure 12. TPC-H SELECT Intensive: Turning On/Off Candidate Selection/Enumeration Techniques.**

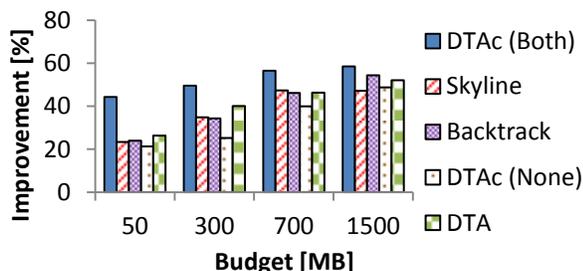

**Figure 13. TPC-H INSERT Intensive: Turning On/Off Candidate Selection/Enumeration Techniques.**

**Candidate Selection and Enumeration**: Second, we verify the effects of techniques we presented in Section 6. Figure 12 and Figure 13 compare the improvements of database designs produced by DTAc and DTA with various space budgets.

*Improvement* is an estimated runtime improvement from the initial database, e.g., improvement of 75% means a 4x speed up while 85% means a 6x speed up. The dataset is TPC-H and we enable only simple indexes (clustered and secondary indexes on tables) this time. The full implementation, denoted as "DTAc (Both)", considers compressed indexes and applies both the Skyline selection and backtracking in the greedy enumeration phase. "Skyline" and "Backtrack", as the name suggests, apply each of them. "DTAc (None)" applies none of them, just generating compressed versions of candidate indexes. "DTA" is the original DTA that does not consider compressed indexes.

As the result shows, only the full implementation achieves significantly better designs especially in tight space budgets. This is because the current candidate selection which picks only the fastest configuration per query cannot capture the potential of compressed indexes with smaller sizes and the current enumeration algorithms cannot choose an index that is slower but saves space. This experiment shows that both the Skyline selection and the greedy backtracking are essential to capture good database designs with compressed indexes.

**Designs for Sales**: Next experiment is on the *Sales* database. We use the same setting as the previous experiment, but this time we run only the full implementation of DTAc. Figure 14 and Figure 15 compare the database designs produced by DTAc and DTA. As the result shows, DTAc achieves significantly better designs because it utilizes compression to make indexes faster and also to allow more indexes within the space budget. Also, DTAc is aware of the overheads of compressed indexes and avoids compressing too many indexes in the INSERT intensive case (all designs for 100MB budget and larger is the same). This prevents the generated design from slowing down with larger budgets, which happens with a naïve design tool that decouples compression from the choice of indexes.

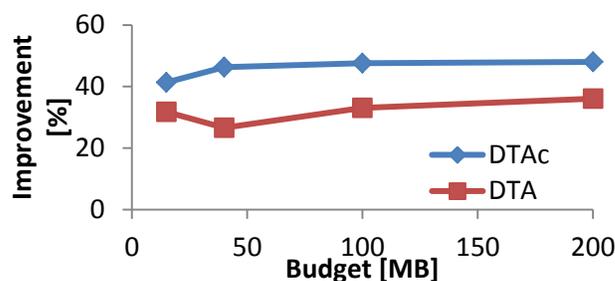

**Figure 14. Sales SELECT Intensive, Simple Indexes.**

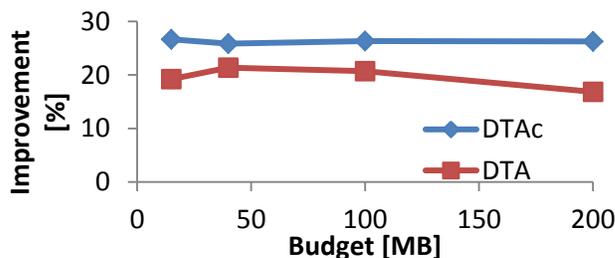

**Figure 15. Sales INSERT Intensive, Simple Indexes.**

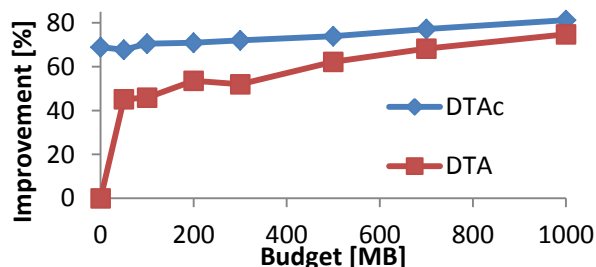

**Figure 16. TPC-H SELECT Intensive, All Features.**

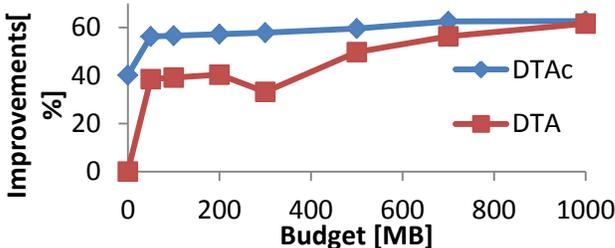

**Figure 17. TPC-H INSERT Intensive, All Features.**

**Designs for TPC-H**: Finally, we show the results with all features (partial indexes and MV indexes too) enabled. Figures 16 and 17 compare DTAc and DTA on TPC-H for SELECT intensive and INSERT intensive workloads. Again, we observe substantially better improvements with DTAc because it utilizes index compression to make the best of limited space. In the SELECT intensive case, the difference is a factor of 2 (e.g., 70% improvement vs. 40% improvement) in tight space budgets. The difference is smaller in larger space budgets (10% to 50%) because more indexes can fit the space budget without compression. In the INSERT intensive case, the designs by DTAc for larger space budgets are similar to DTA since the update overhead of compressed indexes is significant and DTAc chooses not select compressed indexes.